\def\arcsec{$^{\prime\prime}$}
\def\sarc{\(\stackrel{{}^{\prime\prime}}{\textstyle .}\)}

\documentstyle[11pt,emulateapj]{article}
\tighten

\begin{document}

\def\sarc{$^{\prime\prime}\!\!.$}
\def\beginrefer{\section*{References}%
\begin{quotation}\mbox{}\par}
\def\refer#1\par{{\setlength{\parindent}{-\leftmargin}\indent#1\par}}
\def\endrefer{\end{quotation}}
\title{CONSTRAINING H$_{0}$ FROM {\sl CHANDRA} OBSERVATIONS OF Q0957+561 }
\submitted{The Astrophysical Journal, accepted}

\author{G. Chartas\altaffilmark{1}, V. Gupta\altaffilmark{1}, G. Garmire\altaffilmark{1},
C. Jones\altaffilmark{2}, E. E. Falco\altaffilmark{3}, I. I. Shapiro\altaffilmark{2}
and F. Tavecchio\altaffilmark{4}}

\altaffiltext{1}{Astronomy and Astrophysics Department, The Pennsylvania
State University, University Park, PA 16802.}

\altaffiltext{2}{Harvard-Smithsonian Center For Astrophysics, Cambridge, MA 02138.}

\altaffiltext{3} {Smithsonian Institution, F. L. Whipple Observatory,
PO Box 97, 670 Mt. Hopkins Road, Amado, AZ 85645; 
falco@cfa.harvard.edu.}

\altaffiltext{4}{Osservatorio Astronomico di Brera, via Brera 28, Milano, I-20121, Italy}

\begin{abstract}
We report the detection of the lens cluster of the 
gravitational lens (GL) system Q0957+561 from
a deep observation with the Advanced CCD Imaging Spectrometer 
on-board the {\sl Chandra X-ray Observatory}.
Intracluster X-ray emission is found to be centered 4{\sarc}3$_{-1.3}^{+1.3}$ east
and 3{\sarc}5$_{-0.6}^{+1.2}$ north of image B, nearer than previous estimates.
Its spectrum can be modeled well
with a thermal plasma model consistent
with the emission originating from a cluster
at a redshift of 0.36.
Our best-fit estimates of the cluster temperature of 
T$_{e}$ = 2.09$_{-0.54}^{+0.83}$~keV (90\% confidence) and
mass distribution of the cluster are used to derive
the convergence parameter $\kappa$, the ratio of the cluster
surface mass density to the critical density required
for lensing. We estimate the convergence parameter 
at the location of the lensed images A and B to be 
${\kappa_A}$ = 0.22$^{+0.14}_{-0.07}$ and
${\kappa_B}$ = 0.21$^{+0.12}_{-0.07}$, respectively (90\% confidence levels). 
The observed cluster center, mass distribution and convergence parameter $\kappa$
provide additional constraints to lens models of this system.
Our new results break a mass-sheet degeneracy in GL models
of this system and provide better constraints of $\sim$ 29\% 
(90\% confidence levels) on the Hubble constant.
We also present results from the detection of the most distant
X-ray jet (z = 1.41) detected to date. The jet extends approximately 
8{\arcsec} NE of image A and three knots are resolved
along the X-ray jet with flux densities decreasing with distance from the core.
The observed radio and optical flux densities of the knots are fitted well
with a synchrotron model and the X-ray emission is modeled well  
with inverse Compton scattering of Cosmic Microwave Background photons
by synchrotron-emitting electrons in the jet.

\end{abstract}

\section{INTRODUCTION}

In an earlier paper based on deep ROSAT observations, we had reported the 3~$\sigma$ detection of X-rays  
from the cluster of galaxies of the gravitational lens system Q0957+561 
(Chartas et al. 1998). This cluster contributes to the lensing 
of a distant z = 1.41 radio loud quasar. The lensed
quasar appears as two images separated by 6{\sarc}17 and denoted as A
(north) and B (south).
An accurate determination of 
the mass distribution of the cluster is essential in reducing the uncertainty of 
the Hubble constant as derived from the application of 
Refsdal's lensing method (Refsdal, 1964a,1964b).
Due to the limited spatial and spectral resolution and low signal-to-noise ratio 
of the previous X-ray observations, only weak constraints could be placed 
on the cluster properties.  
In particular, we obtained estimates for the convergence parameter, $\kappa$,
the ratio of the projected two-dimensional (2D) surface mass density of the cluster
to the critical surface mass density of the lens cluster.
We found $\kappa$ to range between 0.07 and 0.21, when
assuming that the cluster center was located 24{\arcsec} away from 
image B, a separation based on optical observations of the galaxy members of the cluster
(Angonin-Willaime et al. 1994). 
Other methods for estimating the convergence parameter $\kappa$
rely on weak-lensing of background galaxies (Fischer et al. 1997),
measurements of the velocity dispersion of the lens galaxy G1 
(Falco et al., 1997; Tonry \& Franx 1999; Romanowsky \& Kochanek 1999)
and measurements of the velocity dispersion of the lens cluster
(Garrett, Walsh, \& Carswell 1992, Angonin-Willaime, Soucail, \& Vanderriest 1994).
The present constraints on $\kappa$ are quite poor
and have probably underestimated the systematic errors due to the 
uncertainty of the location of the cluster center and of the mass profile. 
Barkana et al. (1999) found a strong dependence of $\kappa$ 
on the assumed cluster mass profile and on the distance of the cluster center  
from the lens galaxy.  

A measurement of the convergence parameter $\kappa$ 
at the location of the lensed images with respect to the cluster center
is needed to break the Hubble constant mass-sheet degeneracy (Falco et al. 1985).
More direct methods of determining the location of the center of mass of the
cluster are provided by measurements of the intracluster 
medium emission in the X-ray band, 
the weak lensing method, and optical measurements
of the member galaxies. 
X-ray measurements with ROSAT did not provide any useful 
constraints on the location of the cluster
center. The weak lensing method, which determines
the mass distribution of the lens cluster from the 
gravitational distortion of the images of background galaxies,
yields a center of mass of the cluster to be located
$\sim$ 18{\arcsec} east and 13{\arcsec} north of G1, with a 1$\sigma$
uncertainty of about 15{\arcsec}. 
Angonin-Willaime et al. (1994) derived a center
for the cluster at 13{\sarc}7 west, 19{\sarc}6 south of
image B by proportionally weighting
galaxies by their R band luminosity.  This
method was found to be very sensitive to the number of galaxies included in the 
calculation. For example, Garrett et al. (1992) 
using a subset of the galaxies had found the 
cluster center at 3{\sarc}5 west and 1{\sarc}2 north of G1. 
Another method for estimating the cluster center
is based on the ``best-fit'' values for the lens 
models that describe this system. In a recent analysis Chae et al. (1999)
model the cluster contribution by a power-law sphere with an extended core. 
They found a distance between the cluster center and the G1 galaxy 
of 9$''$$_{-4}^{+4}$ and a position angle $\theta$ 
(North through East) of $\sim$ 52$^{\circ}$.
In an independent analysis Barkana et al. (1999) model the cluster 
as a singular isothermal sphere and allow the center position and velocity dispersion
of the cluster to be free parameters. They found a 
value for the distance of the cluster center 
from the G1 galaxy of 13{\sarc}7 east, 6{\sarc}9 north 
and a value for the velocity dispersion of 439~km s$^{-1}$.
The large systematic differences between independent
estimates of the center position of the cluster obtained through 
modeling imply that the mass distribution of the 
lens is not accounted for correctly in these models.
A recent discovery of the lensed images of the host galaxy of the
quasar Q0957+561 by Keeton et al. (2000) has provided tighter
constraints on the global structure of the lensing potential.  
Gravitational-lens (GL) models of Q0957+561 developed prior to the discovery of the host galaxy 
failed to reproduce the correct shape for the host galaxy arcs 
and therefore any constraints they provide on the value of the center location of the cluster
are unreliable. Lens models by Keeton et al. (2000) that incorporate the
constraints provided by the host galaxy arcs
imply that the cluster potential is approximately centered on G1.

In this paper we present results from the recent observation of the 
GL system Q0957+561 
with the {\sl Chandra X-ray Observatory}. Section 2 contains a 
description of the observation and the data reduction.
In section 3 we perform a detailed spatial analysis of 
the X-ray image of the lens cluster and present a more direct measurement
of the location of the cluster center and the cluster core radius.
In section 3 we also briefly describe the detection of an X-ray jet
corresponding to the radio jet from image A. The spectral analyses 
of the lens cluster, lensed images,
and X-ray jet are presented in section 4. We revisit the
determination of the convergence parameter, $\kappa$, in section 5.
In section 6 we discuss the implications of the {\sl Chandra} constraints 
on the Hubble constant and conclude with a summary of 
the results obtained from the {\sl Chandra} observation.
We use $H_{0}$ = 75 km s$^{-1}$ Mpc$^{-1}$,
$q_{0}$ = 0.5, and $\Lambda$ = 0, unless mentioned otherwise.
At the redshifts of 0.36(lens) and 1.41(quasar) an angular size of 1{\arcsec}
corresponds to length scales of $\sim$ 4.1~kpc and $\sim$ 5.7~kpc, respectively.

\section{OBSERVATIONS AND ANALYSIS}
Q0957+561 was observed with the ACIS instrument (Garmire et al. 2001, in 
preparation) onboard the {\sl Chandra X-ray Observatory} in 2000 April 4 for 47,660s.
The data were collected on the back-illuminated
S3 CCD of ACIS. The instrument was operated in sub-array mode where a
restricted region with 254 of 1024 rows of the chip collects data
to reduce the frame-time from the nominal 3.24~s(full-frame) to 0.741~s.   
The telescope pointing was set approximately 1 arcmin off axis.
We used the sub-array mode and off-axis pointing
to reduce the probability of multiple X-rays landing
within a few CCD pixels during the same CCD frame readout.
This effect, commonly referred to as pile-up, may lead to spectral and spatial distortions
and loss of detected events.
The detected count rates for images A and B are 
0.21~counts s$^{-1}$ and 0.16~counts s$^{-1}$, with respective 
estimated pile-up fractions of about 11\% and 7\% for the 0.741~s
frame-time.
We applied the data processing techniques recommended by the 
{\sl Chandra X-ray Center} (CXC) which include
removal of hot pixels and non-X-ray events.
The standard CXC software pipelines randomize the event positions
by $\pm$ 0{\sarc}246 (1 ACIS pixel = 0{\sarc}492), in each
spatial coordinate. This randomization is done to
avoid aliasing effects noticeable in observations
with exposure times less than $\sim$ 2~ks. 
Since the exposure time for the {\sl Chandra} observation of Q0957+561 is
considerably longer than 2~ks, we reprocessed the data with the CXC software tool
\verb+ACISPROCESSEVENTS+ without randomizing
the event positions. This reprocessing led to an improvement of the 
50 percent encircled energy radius from 0{\sarc}40 to 0{\sarc}35.
The CCD read-out streaks of the bright images A and B of Q0957+561 
are removed in the spatial and spectral analysis. These
streaks are produced from events that are recorded on the
CCD during the transfer of charge into the CCD frame-store region.
The background level was fairly constant throughout the
observation with a mean value of 2.2 $\times$ 10$^{-6}$ events s$^{-1}$ arcsec$^{-2}$
in the 0.3-7~keV band. \\

\section{SPATIAL ANALYSIS; NEW X-RAY COMPONENTS IDENTIFIED}

\subsection{The Lens Cluster}

In Figure 1 we show the {\sl Chandra} image of the 
GL system Q0957+561. To enhance the presence of soft extended
X-ray emission, the image was filtered to include  
only photons with energies ranging between  0.5 and 3.0~keV.
This choice of energy range is supported by the analysis of the
extracted spectrum of the cluster (see section 4.1)
which indicates that the cluster spectrum is very soft with most 
of the detected counts lying below 3~keV.
The image has been binned with a bin size of 0{\sarc}5
and smoothed with the software tool \verb+CSMOOTH+
developed by Ebeling et al. (2000) and provided by the 
{\sl Chandra X-ray Center} (CXC).
\verb+CSMOOTH+ smoothes a two-dimensional image with a circular 
Gaussian kernel of varying radius. 
Extended X-ray emission with a radius 
of $\sim$ 30{\arcsec}, encompassing the quasar images,
is clearly visible in Figure 1. The extended emission appears to be centered
slightly northeast of image B. This is seen if one takes
the midpoint of the two outer contour levels of the X-ray emission 
shown in Figure 1. The midpoint is slightly to the east of image B.  

To constrain the morphology of the soft extended emission 
we performed a 2D fit to the X-ray data using the model
described below. The spatial model has the following three components:
extended cluster emission, unresolved emission from the
quasar images and background emission.

(a) To describe the cluster brightness profile,
we use a $\beta$ model

\begin{eqnarray}
{\centering f(r) = A[1 + (\frac{r}{r_c})^{2}]^{-3{\beta} + \frac{1}{2}}}
\end{eqnarray}

\noindent
where,  

\begin{eqnarray}
r(x,y)~&=&~\frac{\sqrt{{x^{\prime}}^2(1-{\epsilon})^2 + {y^{\prime}}^2}}{1-{\epsilon}} \nonumber \\
x^{\prime}~&=&~(x - x_{\rm0})\cos({\theta}) + (y - y_{\rm0})\sin({\theta}) \nonumber \\
y^{\prime}~&=&~(y - y_{\rm0})\cos({\theta}) - (x - x_{\rm0})\sin({\theta}) \,, \nonumber
\end{eqnarray}

\noindent
$\epsilon$ is the ellipticity of the model 
defined here as $\epsilon = 1-q$; $q$ is the axis ratio;
$\theta$ is defined as the angle between the major axis 
and west, and is measured west to north;
$x_{\rm0}, y_{\rm0}$ are the positions of the cluster center; 
and $\beta$ is the ratio of kinetic energy per unit mass
in galaxies to kinetic energy per unit mass in gas.

(b) To model the lensed quasar images, we use
simulated point spread functions (PSF's). The centroids of the
model PSF's are fixed to the observed image centroids.
The relative normalization of the PSFs is 
the flux ratio of the lensed images as determined 
from annular regions of inner and outer radii of  0{\sarc}5 and 2{\arcsec},
respectively, centered on the images. These annuli were
chosen to avoid the slightly piled-up cores of the images.
The model PSFs appropriate for this 
observation were created employing the simulation tool \verb+MARX v3+
(Wise et al. 1997) with an input spectrum 
derived from the best-fit {\sl Chandra} spectrum of the lensed images. 
Specifically, we used an absorbed power law with a
column density of N$_{H}$ = 0.82 $\times$ 10$^{20}$ cm$^{-2}$ (Dickey \& Lockman, 1990)
and photon indices of $\Gamma$ = 2.08 and 1.94 for images A and B respectively (see section 4.2).
Here-after the term best-fit is used to describe a 
result obtained by fitting a model to 
our data using an optimization technique
to find the local fit-statistic minimum.
The Levenberg-Marquardt optimization method 
is used for the spectral analysis, 
the Powell method is used for the spatial analysis of the cluster and the 
downhill simplex method is used by the fitting tool \verb+LYNX+ (section 4.2).

(c) Finally in our model, we include a uniform background of 
0.01 events per pixel obtained from a background region at 
the same distance from the aim point as the cluster. 

In determining the cluster properties, we
omitted a 10{\arcsec} radius region centered on the midpoint between the quasar
images to avoid biasing the fit due to residuals
in modeling the cores of the PSFs. 
For the annulus from 8{\arcsec} to 40{\arcsec} centered on the midpoint of
the quasar images, we binned the image in 1{\arcsec} pixels and smoothed this
with a Gaussian ($\sigma = 3{\arcsec}$) prior to performing the spatial
fitting. The fits were performed with the CXC software package \verb+SHERPA+. 
To evaluate the sensitivity of the results to the choice of 
assumptions made for sizes of spatial windows
and bin sizes of the data, we performed the spatial 
analysis over a wide range of input model parameters.
Specifically, we binned the X-ray data with bin sizes
varying between 1{\arcsec} and 2{\arcsec}. 
The annulus region used to model the cluster 
was varied with an inner radius ranging between
10{\arcsec} and 14{\arcsec} and an outer radius ranging
between 30{\arcsec} and 40{\arcsec}.
The errors quoted for the spatial model parameters
represent the maximum uncertainties  
obtained from all spatial fits. 
Results for our spatial fits are presented in Table 1.
We find the center of the mass of the lens cluster to be
located at ${\Delta}{\alpha}$ = 4{\sarc}3$_{-1.3}^{+1.3}$ east, 
${\Delta}{\delta}$ = 3{\sarc}5$_{-0.6}^{+1.2}$ north of the
core of image B. The fits indicate that the smoothed mass distribution
of the cluster is close to spherical
with an ellipticity ${\epsilon} = 0.19_{-0.08}^{+0.06}$. 
The best-fit values for $\beta$ and the core radius of the cluster
are $\beta$ = 0.47$_{-0.06}^{+0.06}$ and r$_{0}$ = 15{\sarc}4$_{-3.5}^{+3.5}$ (62.5 kpc),
respectively. Our values for $\beta$ and $r_{0}$, within
the uncertainties, fall on the trend line of $\beta$ versus
$r_{0}$ obtained previously for a suite of galaxy clusters (Jones \& Forman 1999).

\subsection{The X-ray Jet of Image A} 

We notice a faint feature in the {\sl Chandra} 
image extending northeast of image A
that is aligned with the radio jet and appears to have a jet-like morphology. The X-ray jet is significant at 
the 4 $\sigma$ level. At a redshift of 1.41 this is the most distant X-ray jet detected 
to date. An overlay between this X-ray jet feature 
and a 3.6~cm radio map of the jet in image A (radio data from Harvanek et al. 1997) 
is shown in Figure 2. The X-ray and radio jet morphologies appear to be somewhat similar.
A comparison between the radio and X-ray jet profiles along the jet ridge line 
is shown in Figure 3. 
The X-ray jet extends NE about 8{\arcsec} from the core of image A 
with bright knots at 2{\sarc}3, 4{\arcsec} and 6{\arcsec} (here-after 
also referred to as knots A, B and C, respectively) from the core.
Each of these knots coincides within 0{\sarc}5 
with radio knots at similar locations.
The intensity of the X-ray knots appears to decay as a 
function of distance from the core 
in contrast to the radio knots which become 
brighter at larger distances from the core.  
One possible interpretation of the former is that the jet flow
is decelerating and the brightness change is
due to aging of the higher energy electron population of the jet.
The increase in the radio brightness along the jet 
may be due to an increase of the magnetic field strength.
A similar anti-correlation between radio and X-ray profiles 
was recently reported in 3C273 (Sambruna et al. 2001; Marshall et al. 2001).
Lens models for Q0957+561 predict a small magnification
gradient along the jet. We estimate that the magnifications 
for the knots B and C are 1.9 and 1.6, respectively.
Therefore, the magnification gradient 
does not greatly change the
X-ray and radio brightness profiles.
In section 4.3 we briefly discuss possible
mechanisms that may explain the origin of the X-ray jet 
emission and present the spectral energy distribution 
for knots B and C.

\section{SPECTRAL ANALYSIS}

\subsection{A Cool Lens Cluster}
The spectrum for the cluster of galaxies was extracted from a 
40{\arcsec}-radius circle centered on the X-ray determined cluster center. 
Beyond this radius, cluster emission is not detected above the background.
Three-arcsecond-radius circles centered on images A and B were excluded.
Emission from the X-ray jet was also excluded by omitting a 
rectangle 2{\sarc}5 by 7{\arcsec}.
However, even beyond the 3{\arcsec} radius,
mirror scattering of the bright lensed images produces a
significant contamination of the cluster spectrum, particularly
at hard energies.
We estimate the spectrum of the contamination produced by images A and B
to the cluster spectrum  by performing  simulations with the 
raytrace simulator tool \verb+MARX+ provided by the CXC.
We simulate two point sources 
centered at the locations of images A and B with input spectra 
and normalizations derived from our spectral analysis of these images. 
We find a total of about $\sim$ 200 counts due to quasar emission 
($\sim$ 0.8\% of the total counts from images
A and B) within the cluster extraction region.
The CCD background and quasar contamination are 
subtracted from the cluster region resulting in the spectrum 
shown in Figure 4. A net total of about $\sim$ 600 X-ray events
originate from the cluster. 
The data were fit with the spectral analysis tool \verb+XSPEC+ (Arnaud 1996).
We model the cluster spectrum with a Raymond - Smith 
thermal plasma modified by absorption due 
to our Galaxy. The  Galactic column density
is fixed at the value of N$_{H}$ = 0.82 $\times$ 10$^{20}$ cm$^{-2}$
for all spectral fits performed in our analysis.  
We assumed a typical value for the metal abundances in clusters of 30\% solar
(see, e.g., Henriksen 1985; Hughes et al. 1988; and Arnaud et al. 1987).
Due to the relatively low signal-to-noise ratio of the
cluster spectrum the metal abundances of the cluster cannot be constrained
within useful limits. If we let the abundance be a free parameter we obtain a
best-fit value for the abundance of A = 0.19$_{-0.19}^{+0.54}$
(90\% confidence level) and a best fit temperature of 2.1keV.
We note that the choice of metal abundance (between
0 and 100\% solar) has little effect on the temperature determinations.
The abundance ratios used in the Raymond - Smith thermal plasma emission
model are those of Anders \& Grevesse (1989).
Our best-fit model is shown in Figure 4. 
The uncertainty in the calibration of ACIS S3 below 0.5~keV
contributes to the large residuals between 0.4 and 0.5~keV.
We find a temperature for the cluster of T$_{e}$ = 2.09$_{-0.54}^{+0.83}$~keV
at the 90\% confidence level. The spectral line features between
0.7 and 0.9~keV (observed frame) correspond  to a redshifted z= 0.36
complex of Fe~L lines. Our spectral analysis thus confirms that 
the extended emission originates from the lens.
The relatively low cluster temperature of $\sim$ 2~keV
and our values of $\beta$ and the core radius from section 3.1,
within their uncertainties, 
are consistent with the observed correlations 
between temperature, $\beta$ and core radius 
obtained previously for a large sample 
of clusters of galaxies (Jones \& Forman 1999).
We estimate that $\sim$ 3\% of the counts originating
from cluster emission are excluded from our
extracted spectrum due to the regions used in our 
analysis.
Correcting for this effect we find a 2 - 10~keV 
cluster luminosity of 4.7 $\times$ 10$^{42}$ erg s$^{-1}$.
Our estimated values for L$_{X}$ and T$_{e}$ are consistent with 
the empirical L$_{X}$ - T$_{e}$ correlation between cluster 
luminosity and temperature of, 
$T_{keV}$ = 0.13(+0.08,-0.07) $\times$ $L_{40}^{(0.351 \pm 0.068)}$,
where $L_{40}$ is the measured 0.5-4.5~keV luminosity
in units of 10$^{40}$ erg s$^{-1}$ ($H_{0}$ = 50 km s$^{-1}$ Mpc$^{-1}$)
(Jones \& Forman 1999; Markevitch, 1998). 
Specifically, the L$_{X}$ - T$_{e}$ correlation
yields a value of T$_{e}$ $\sim$ 2~keV
for our value of $L_{40}$ = 2.3 $\times$ 10$^{3}$.

\subsection{The Spectra of Images A and B}

As mentioned in section 2, we expect the spectra of images A 
and B to be slightly piled-up. 
We used two independent methods to account for pile-up.

In the first method we extracted spectra
from annuli centered on each image with inner and outer radii
of 0{\sarc}5 and 3{\arcsec}, respectively. 
Pile-up is significantly reduced beyond the 0{\sarc}5 radius, in the wings of the PSFs.
One of the drawbacks of the annulus method is that only 
$\sim$ 40\% of the total quasar counts are considered in the spectral fit, thus, leading to 
larger uncertainties in the values of the estimated parameters.
We corrected the ancillary files provided by the CXC to account for the energy dependence
of X-ray scattering in the {\sl Chandra} mirrors 
within the selected annuli.
The energy dependent correction function was evaluated by 
performing raytrace simulations with the software 
tool \verb+MARX+. We modeled the spectra with power-laws 
modified by absorption from our Galaxy.
The fits are statistically acceptable ($\chi^{2}(\nu) = 113(138)$ 
and $\chi^{2}(\nu) =109(105)$, for images A and B,
where $\nu$ is the number of degrees of freedom) 
with best-fit values for the photon indices of images A and B of
2.06$^{+0.04}_{-0.05}$ and 2.06$^{+0.05}_{-0.05}$,
respectively (90\% confidence errors). 
The flux ratio B/A is 0.74 $\pm$ 0.02, consistent with the
observed flux ratio in the radio band of 0.76 $\pm$ 0.03 (VLBI $\lambda$ 13cm,
Falco et al. 1991) and 0.72 $\pm$ 0.04 (VLA $\lambda$ 6cm core, Conner et al. 1992).
Previous measurements of the X-ray flux ratios 
of 0.3 $\pm$ 0.1 and 1.5 $\pm$ 0.2 made
with {\sl EINSTEIN} and {\sl ROSAT}, respectively,(Chartas et al. 1998)
differed significantly from this value, suggesting
the presence of microlensing for those epochs.

For our second approach we used the forward fitting tool \verb+LYNX+
developed at PSU (Chartas et al. 2000). 
Spectra were extracted from circles centered on each image with 
radii of 3{\arcsec}.
We found the photon indices 
for A and B to be 2.08 $\pm$ 0.03 and 1.94 $\pm$ 0.03 (90\% confidence errors).
The 0.5-10~keV X-ray fluxes of images A and B corrected for pile-up 
are 11.6 $\pm$ 0.9 $\times$ 10$^{-13}$ erg s$^{-1}$ cm$^{-2}$ and 
8.8 $\pm$ 0.7 $\times$ 10$^{-13}$ erg s$^{-1}$ cm$^{-2}$. 
The errors in the estimates of the photon indices
in both methods do not include systematic errors due to
the uncertainties in the ACIS detector quantum efficiency
and energy responce. We note that the ``annulus'' method uses
detector response matrices provided by the CXC
whereas \verb+LYNX+ links to a Monte Carlo simulator of ACIS developed by the ACIS
instrument team (Townsley et al., in preparation).
We do not detect any line features in their spectra.  
Combining the spectra of images A and B, we place a 95\% confidence 
upper limit of $\sim$ 60~eV (observed-frame) 
on the equivalent width of an intrinsically narrow 
fluorescent Fe~K${\alpha}$ line at 6.4~keV (2.66~keV observed-frame).

\subsection{The Spectral Energy Distribution of the Jet}
In Figure 5 we show the spectral energy distribution(SED) of the two brightest knots B and C
of the jet. The radio flux density, $F_{\nu}$, at 3.6~cm for jet A was based on  
the value reported by Harvanek et al. (1997).
The jet knots have not been detected in the optical band;
we therefore placed upper limits on the optical flux density 
of the knots based on deep HST observations
of this field performed by Bernstein \& Fischer (1997).
We find the 3$\sigma$ upper limit on ${\nu}F_{\nu}$ 
at 5500 {\AA}~to be 5.6 $\times$ 10$^{-16}$ erg s$^{-1}$ cm$^{-2}$.   
The X-ray spectrum was extracted from a 
2.5{\arcsec} by 7{\arcsec} rectangle. The background was 
extracted from several similar rectangles
located at the same distance from image A but at different 
azimuths. A net total of 50 $\pm$ 8 X-ray events were 
those ascribed to the jet.
We modeled the spectrum with a power-law 
modified by absorption due to our Galaxy.
We find a best-fit value for the photon index 
$\Gamma$ of 1.87 $\pm$ 0.6 (90\% confidence errors) including systematic errors estimated 
from using different background regions.
We find ${\nu}F_{\nu}$ at 1~keV for knots B and C of
8.9 $\pm$ 2.8 $\times$ 10$^{-16}$ and 2.7 $\pm$ 0.9 $\times$ 10$^{-16}$ erg s$^{-1}$ cm$^{-2}$, 
respectively (90\% confidence errors). A simple synchrotron model with a single power law 
distribution is not consistent with the SED of the jet in Q0957A.
Given the high redshift of the jet, the energy density
of the Cosmic Microwave Background (CMB) is enhanced
by a factor of $(1 + z)^4 = 33.7$ over
the local value. We therefore modeled the spectra of knots B and C
with a CMB model, in which cosmic microwave photons are inverse
Compton scattered by synchrotron-emitting electrons in the jet.
The solid and dashed lines in Figure 5 are the fits
of the CMB model to the SEDs of knots B and C, respectively.
We are able to fit both knots well with very moderate
Doppler beaming (see Figure 5).
We conclude that a plausible mechanism to 
explain the X-ray emission is inverse Compton scattering of CMB photons. 
Such a mechanism was shown to be consistent with the peculiar SED
of the jet in PKS~0637-752 (Tavecchio et al, 2000).

\section{MASS DISTRIBUTION AND CONVERGENCE OF THE LENS CLUSTER}

For isothermal and spherical clusters of galaxies 
the equation for hydrostatic equilibrium can be solved for 
the virial mass, $M_{grav}(<r)$, within a radius $r$,

\begin{equation}
{M_{grav}(<r)}  = -{{{k T r}\over{\mu m_{p} G}}{{d~ln~\rho}\over{ d~ln~r}}}
\end{equation}

\noindent 
where $k$ is Boltzmann's constant, $\mu$$m_{p}$ is the mean molecular weight of the cluster gas,
and $\rho(r)$ is the cluster gas density at $r$.

Assuming a $\beta$ model for the density profile of the hot gas
(e.g., Jones \& Forman 1984) we obtain the equation for the total mass of the lens within a radius $r$,

\begin{equation}
{M_{grav}(<r)}  = { {{3 \beta k T}\over{\mu m_{p} G}} { {r} \over {[1 + ({{r_c}\over{r}})^2]}}} {\  ,}
\end{equation}

By incorporating the best-fit spatial and spectral parameters
from our present analysis (see Tables 1 and 2) we find the 
total cluster mass within a radius 
of 1~$h_{75}$$^{-1}$~Mpc to be 
M$_{grav}$ = 9.9$_{-3.8}^{+1.9}$ $\times$ 10$^{13}$ M$_{\odot}$. 
In Figure 6 we show the total cluster mass within a radius $r$
as a function of radius. The shaded region indicates the allowed
values for the cluster mass including the uncertainities
obtained from the spatial and spectral fits to the lens cluster (see Tables 1 and 2).

These mass estimates were used to evaluate the convergence
parameter ${\kappa}(x)$,

\begin{displaymath}
{\kappa(x) = {{\Sigma(x)}\over{\Sigma_{cr}}}} \,
\end{displaymath}
where ${\Sigma(x)}$ is the surface mass density of the lens cluster 
as a function of the cylindrical radius $x$ (Chartas et al. 1998) and
$\Sigma_{cr}$ is the critical surface mass density 
(see, e.g., Schneider, Ehlers \& Falco 1992),
In Figure 7 we plot the convergence parameter $\kappa(x)$
as a function of distance from the cluster center.
The thick solid line corresponds to the best-fit spatial and spectral parameters.
The largest contributor to the uncertainty in our estimate 
of $\kappa(x)$ is the weak constraint on the temperature of the cluster.
To illustrate this we have plotted the uncertainty in
$\kappa(x)$ assuming 68\% (dotted lines) and 90\% (dashed lines) 
confidence intervals for the temperature. 
We also chose cluster limits ranging from 0.7$r_{500}$ and 1.4$r_{500}$,
where $r_{500}$ is the radius in which the mean over-density is 500,
and $r_{500}$ = 1.58(T$_{X}$/10~keV)$^{1/2}$~$h^{-1}_{75}$ Mpc $\sim$ 0.71 $h^{-1}_{75}$ Mpc
(Mohr, Mathiesen, \& Evrard, 1999).
The uncertainity in $\kappa$ introduced by assuming
the cluster limit to range between 0.7 and 1.4$r_{500}$ is
only $\sim$ 2\%. Our assumption of spherical symmetry of the mass distribution
has little effect on the estimate of the convergence parameter.
Calculations of the mass distribution of the nonsperical cluster of
galaxies A2256 (axis ratio $\sim$ 1.6) by Fabricant et al. (1984) showed that
radially integrated mass estimates were negligibly
affected by including an oblate or prolate cluster geometry.
At the best-fit location of images A and B with respect to the cluster center, 
we estimate the convergences assuming the 90\% confidence range in cluster temperature
and find ${\kappa_A}$ = 0.22$^{+0.14}_{-0.07}$ and 
${\kappa_B}$ = 0.21$^{+0.12}_{-0.07}$, respectively.
Our estimated range of uncertainty in $\kappa$ does not include
possible systematic errors arising from the estimation
of the the surface mass density of the galaxy cluster through the use
of a hydrostatic, isothermal $\beta$ model.
Evrard, Metzler \& Navarro (1999) have
performed simulations to investigate the accuracy of
galaxy cluster mass estimates based on X-ray observations.
They found that estimates based on the hydrostatic, isothermal $\beta$ model
are unbiased estimates of the mass of relaxed clusters
with standard deviations of less than 30\%.
Recent observations of clusters of galaxies with XMM-Newton
indicate that the cluster temperature profiles are remarkably isothermal beyond the
central cooling flow (e.g., Arnaud et al. 2001).
We therefore expect that our estimated uncertainity in $\kappa$
is dominated by the large uncertainity in the estimated
value of the cluster temperature and not systematic errors
introduced from the use of the isothermal $\beta$ model.

\section{DISCUSSION AND CONCLUDIONS}
Detailed lens models of Q0957+561 include the cluster's
contribution to the lensing potential by expanding the potential of
the cluster in a Taylor series and keeping terms of up to third order
(Kochanek 1991, Bernstein \& Fischer 1999, Keeton et al. 2000).  The
second order term of the expansion represents the shear from the
cluster and can be expressed as 
${\gamma} = {{\kappa}\over{({1 +{\beta_{rd}}^2})^{3/2}}}$ 
where, ${\beta_{rd}} = r_{c}/d_{c}$, $r_{c}$
is the cluster core radius and $d_{c}$ is the distance from the
cluster to the galaxy G1 (see, Kochanek, 1991). Recent models of
Q0957+561 that incorporate additional constraints from observed arcs
produced by the lensing of the quasar host galaxy imply a cluster
shear amplitude, $\gamma$, that is relatively small compared to the
convergence parameter $\kappa$ (Keeton et al. 2000).  Other
observations of Q0957+561 also favor values of $\kappa$ $ > > $
$\gamma$ (Fischer et al. 1997; Romanowsky \& Kochanek 1999; Chartas et
al. 1998).  As pointed out by Keeton et al. (2000), one way of
producing $\gamma$ $ < < $ $\kappa$ is to have the distance from the
center of the cluster to the center of the lens galaxy be: less than
the core radius for a cluster with a singular isothermal ellipsoid
potential, or: less than the ``universal'' scale length for a cluster
potential that follows the dark matter profile of Navarro, Frenk, and
White (1996).  Measurements at that time, ie., the late 1990s,
indicated that the cluster center was located several core radii away
from the lens (Fischer et al. 1997; Angonin-Willaime et al. 1994) and
therefore were not consistent with the large value of $\kappa$
relative to the amplitude of the shear.

The {\sl Chandra} observation of the cluster resolves this apparent
discrepancy. We find the distance between the center of mass of the
cluster and the center of the lens galaxy G1 to be $d_{c}$ = 4{\sarc}8
considerably smaller than suggested from previous estimates.  The
best-fit position angle (North through East) and core radius of the
cluster are $\sim$ 59$^{\circ}$ and 15{\sarc}4$^{+3.5}_{-3.5}$,
respectively.  The {\sl Chandra} observations of Q0957+561 therefore
indicate that the cluster center is located within a core radius of
the lens galaxy G1.  Using the values for $r_{c}$, $d_{c}$, and
${\kappa}$ provided by our analysis of the {\sl Chandra} observation of
Q0957+561 we find that the cluster shear amplitude is $\gamma$ = 0.01
$\pm$ 0.009 (90\% confidence level), consistent with the recent model
results of Keeton et al. (2000).
The {\sl Chandra} observation of Q0957+561 has eliminated 
several uncertainties introduced in our previous analysis of the ROSAT 
HRI data of this system. Specifically, the temperature, the core radius, 
and cluster shape are now determined more reliably. 
Previous estimates of $\kappa$ are unreliable due to the large 
distances between the center of the cluster and the galaxy G1
assumed in these analyses. 
To evaluate the implication of our estimate
of $\kappa$ on the Hubble constant, we write 
$H_{0}$ = $H'_{0}$(1 - $\overline{\kappa}$) = 100$h$ km s$^{-1}$ Mpc$^{-1}$ 
,where we define the average convergence parameter 
of the cluster at the image locations as  
$\overline{\kappa} = ({\kappa}_{A} + {\kappa_B})/2$.
Due to the mass-sheet degeneracy problem, lens models provide 
only the model-dependent value $h$/(1 - $\overline{\kappa}$).
Our present observation of the lens cluster  
results in a $\sim$ $\pm$ 14\%  uncertainty in 
the quantity (1 - $\overline{\kappa}$).
The resulting uncertainty in $H_{0}$ is 
$\sim$ $\pm$ 29\% , where we have assumed an uncertainty of $\sim$ $\pm$ 25\% 
in $H'_{0}$ based on the analysis of Keeton et al. (2000). 
We anticipate that the next generation of lens models for Q0957+561
may provide tighter constraints on $H'_{0}$ when
the constraints on the location, ellipticity and convergence of 
the cluster, based on the {\sl Chandra} observations, are incorporated. 

As described in section 5 the largest contributor to the present uncertainty in
$\kappa$ obtained by the X-ray method is the temperature of the intracluster gas.
Given that the cluster is located near the quasar images, we
may optimize a future observation of Q0957 by 
moving the telescope aim point closer to the cluster center 
to improve the spatial resolution 
and reduce the contamination of the 
cluster by the bright images. This change will reduce the uncertainty
of the estimates of the spectral and spatial parameter values for the cluster.

We would like to thank Daniel Harris and William Forman for helpful discussions.
We acknowledge financial support by NASA grant NAS 8-38252
and support from the Smithsonian Institution.

\newpage
\normalsize
\begin{center}
\begin{tabular}{ccccc}
\multicolumn{5}{c}{TABLE 1}\\
\multicolumn{5}{c}{Results From Spatial Fits To Lens Cluster} \\
& & & &  \\ \hline\hline
\multicolumn{1}{c} {$x_{0}$$^{a}$} &
\multicolumn{1}{c} {$y_{0}$$^{a}$} &
\multicolumn{1}{c} {$\beta$$^{b}$} &
\multicolumn{1}{c} {$r_{0}$$^{c}$} &
\multicolumn{1}{c} {${\epsilon}^{d}$} \\
         &             &       &        &                          \\
  ({\arcsec})   &  ({\arcsec})    &       &  ({\arcsec}) &              \\ \hline
& & & & \\
4.3$_{-1.3}^{+1.3}$ & 3.5$_{-0.6}^{+1.2}$ & 0.47$_{-0.06}^{+0.06}$ & 15.4$_{-3.5}^{+3.5}$ & 0.19$_{-0.08}^{+0.06}$    \\
& & & & \\
\hline \hline
\end{tabular}
\end{center}

NOTES-\\ The probability distributions for the best-fit model parameters
$y_{0}$ and ${\epsilon}$ derived from our error analysis are not Gaussian and 
any value of $y_{0}$ and ${\epsilon}$ within the quoted errors should be considered to have similar likelihood. The errors quoted for the spatial model parameters, 
represent the maximum range of the parameter values obtained from 
the suite of spatial fits performed in our sensitivity analysis.\\ 
$^{a}$ $x_{0}$ and $y_{0}$ are the separations ${\Delta}{\alpha}$ 
(a positive value indicates east of B) and ${\Delta}{\delta}$ (a positive value indicates north
of B) of the center of the lens cluster from the core of image B. \\
$^{b}$ $\beta$ is the ratio of kinetic energy per unit mass
in galaxies to kinetic energy per unit mass in the gas. 
$\beta$ is determined from the fit to the surface brightness of the lens cluster.\\
$^{c}$ $r_{0}$ is the core radius of the lens cluster. \\
$^{d}$ ${\epsilon}$ is the ellipticity of the cluster defined as ${\epsilon}$ = 1 - $q$,
where $q$ is the axis ratio.\\

\newpage
\normalsize
\begin{center}
\begin{tabular}{cccccc}
\multicolumn{6}{c}{TABLE 2}\\
\multicolumn{6}{c}{Values of Model Parameters Determined from fits to the Spectrum of the Lens Cluster} \\
& & & & & \\ \hline\hline
\multicolumn{1}{c} {Model} &
\multicolumn{1}{c} {N$_{H}$ } &
\multicolumn{1}{c} {T$_{e}$} &
\multicolumn{1}{c} {F$_{X}$} &
\multicolumn{1}{c} {L$_{X}$}  &
\multicolumn{1}{c} {$\chi^{2}_{\nu}({\nu})$} \\ 
          & cm$^{-2}$                           & keV                    & erg s$^{-1}$ cm$^{-2}$   & erg s$^{-1}$           &    \\ \hline
 &&&&& \\
ABS + RS &0.82 $\times$ 10$^{20}$&  2.09$_{-0.54}^{+0.83}$   & 1.1$\times$ 10$^{-14}$ & 4.7 $\times$ 10$^{42}$&1.3(43)   \\
&&&&& \\
\hline \hline
\end{tabular}
\end{center}

NOTES-\\ The spectra are described by a thermal Raymond - Smith 
model plus absorption due to
cold material at solar abundances fixed to the Galactic value.
The absorbed flux F$_{X}$ and unabsorbed luminosity  L$_{X}$ of the lens cluster
are estimated for the energy range between 2 and 10~keV.  
The reduced chi-squared is defined as $\chi^{2}_{\nu}$ = $\chi^{2}/{\nu}$, where ${\nu}$, 
the number of degrees of freedom, is given in parentheses.

\clearpage

\beginrefer
\refer Anders, E. \& Grevesse, N. 1989, Geochimica et Cosmochimica Acta, 53, 197\\

\refer Angonin-Willaimem M. C., Soucail, G., \& Vanderriest, C. 1994, A\&A, 291, 411 \\

\refer Arnaud, M., Neumann, D.~M., Aghanim, N., Gastaud, R., 
Majerowicz, S., \& Hughes, J.~P., 2001, \aap, 365, L80 \\

\refer Arnaud, K., Johnstone, R., Fabian, A., Crawford, C., Nulsen, P., Shafer, R.,
and Mushotzky, R. 1987, MNRAS., 227, 241. \\

\refer Arnaud, K. A. 1996, ASP Conf. Ser. 101:
Astronomical Data Analysis Software and Systems V, ed. G. Jacoby \& J. Barnes (San Francisco: ASP), 17\\

\refer Barkana, R., Leh{\'a}r, J., Falco, E. E., Grogin, N. A., Keeton, C. R.,
\& Shapiro, I. I. 1999, ApJ, 523, 54\\

\refer Bernstein, G. \& Fischer, P. 1999, \aj, 118, 14 \\

\refer Blandford, R. D. \& Narayan, R. \araa, 1992, 30, 311 \\

\refer Chae, K.-H. 1999, ApJ, 524, 582 \\

\refer Chartas, G., Chuss, D., Forman, W., Jones, C., \& Shapiro, I. I. 1998, ApJ, 504, 661 \\

\refer Chartas, G., Worrall, D. M., Birkinshaw, M., 
Cresitello-Dittmar, M., Cui, W., Ghosh, K. K., 
Harris, D. E., Hooper, E. J., Jauncey, D. L., 
Kim, D. - W., Lovell, J., Mathur, S., Schwartz, D. A., 
Tingay, S. J., Virani, S. N., \& Wilkes, B. J., 2000, \apj, 542, 655 \\

\refer Conner, S. R., and Lehar, J., \& Burke, B. F. 1992, \apj, 387, L61 \\

\refer Dickey, J. M., \& Lockman, F. J., 1990, Ann. Rev. Ast. Astr. 28, 215 \\

\refer Ebeling, H., White, D. A., Rangarajan F.V.N. 2000, MNRAS, submitted \\

\refer Fabricant, D., Rybicki, G., \& Gorenstein, P., 1984, \apj, 286, 186 \\

\refer Falco, E. E., Gorenstein, M. V., \& Shapiro, I. I. 1985, \apj, 289, L1 \\

\refer Falco, E. E., Gorenstein, M. V., \& Shapiro, I. I. 1991, \apj, 372, 364 \\

\refer Falco, E.\ E., Shapiro, I.\ I., Moustakas, L.\ A., \& Davis, M.\ 1997, 
\apj, 484, 70 \\

\refer Fischer, P., Bernstein, G., Rhee, G., \& Tyson, J. A. 1997, A\&A, 113, 521 \\

\refer Garrett, M., Walsh, D., \& Carswell, R. 1992, MNRAS 254, 27 \\

\refer Harvanek, M., Stocke, J. T., Morse, J. A., \& 
Rhee, G. 1997, \aj, 114, 2240 \\

\refer Henriksen, M. 1985 Ph.D. thesis, University of Maryland.\\

\refer Hughes, J., Yamashita, K., Okumura, Y., Tsunemi, H., and Matsuoka, M., 1988, ApJ, 327, 615. \\

\refer Jones. C., \& Forman, W. 1984, \apj, 276, 38\\

\refer Jones. C., \& Forman, W. 1999, \apj, 511, 65\\

\refer Keeton, C. R., Falco, E. E., Impey, C. D., Kochanek, C. S., Leh{\'a}r, J.,
McLeod, B. A., Rix, H.-W., Mu{\~n}oz, J. A., \& Peng, C. Y. 2000, ApJ, 542, 74 \\

\refer Kochanek, 1991, \apj, 382, 58\\

\refer Markevitch, M, 1998, \apj, 504, 27\\

\refer Marshall, H. L., Harris, D. E., Grimes, J. P., 
Drake, J. J., Fruscione, A., Juda, M., Kraft, R. P., 
Mathur, S., Murray, S. S., Ogle, P. M., Pease, D. O., 
Schwartz, D. A., Siemiginowska, A. L. Vrtilek, S. D., 
\& Wargelin, B. J. 2001, \apj, 549, L167 \\

\refer Mohr, J. J., Mathiesen, B. \& Evrard, A. E., 1999, \apj, 517, 627 \\

\refer Navarro, J. F., Frenk, C. S., \& White, S. D. M. 1996, ApJ, 462, 563 \\

\refer Refsdal, S. 1964, MNRAS, 128, 295 \\

\refer Refsdal, S. 1964, MNRAS, 128, 307 \\

\refer Romanowsky, A. J., \& Kochanek, C. S. 1999, ApJ, 516, 18 \\

\refer Sambruna, R., M., Urry, C., M., Tavecchio, F., 
Maraschi, L., Scarpa, R., Chartas, G. \& Muxlow, T. 2001, \apj, 549, L161 \\

\refer Schneider, P., Ehlers, J., \& Falco, E. E., 1992, Gravitational Lensing (New York: Springer) \\

\refer Tavecchio, F., Maraschi, L., Sambruna, R. M., 
Urry, C. M. 2000, 544, L23 \\

\refer Tonry, J., L., \& Franx, M. 1999, \apj, 515, 512 \\

\refer Townsley, L. K., Broos, P. S., Chartas, G., Moskalenko, E.,
Nousek, J. A., \& Pavlov, G.G. in preparation\\

\refer Wise, M. W., Davis, J. E., Huenemoerder, Houck, J. C., Dewey, D.
Flanagan, K. A., and Baluta, C. 1997,
{\it The MARX 3.0 User Guide, CXC Internal Document}
available at http://space.mit.edu/ASC/MARX/\\

\endrefer

\clearpage
\begin{figure*}[t]
\plotfiddle{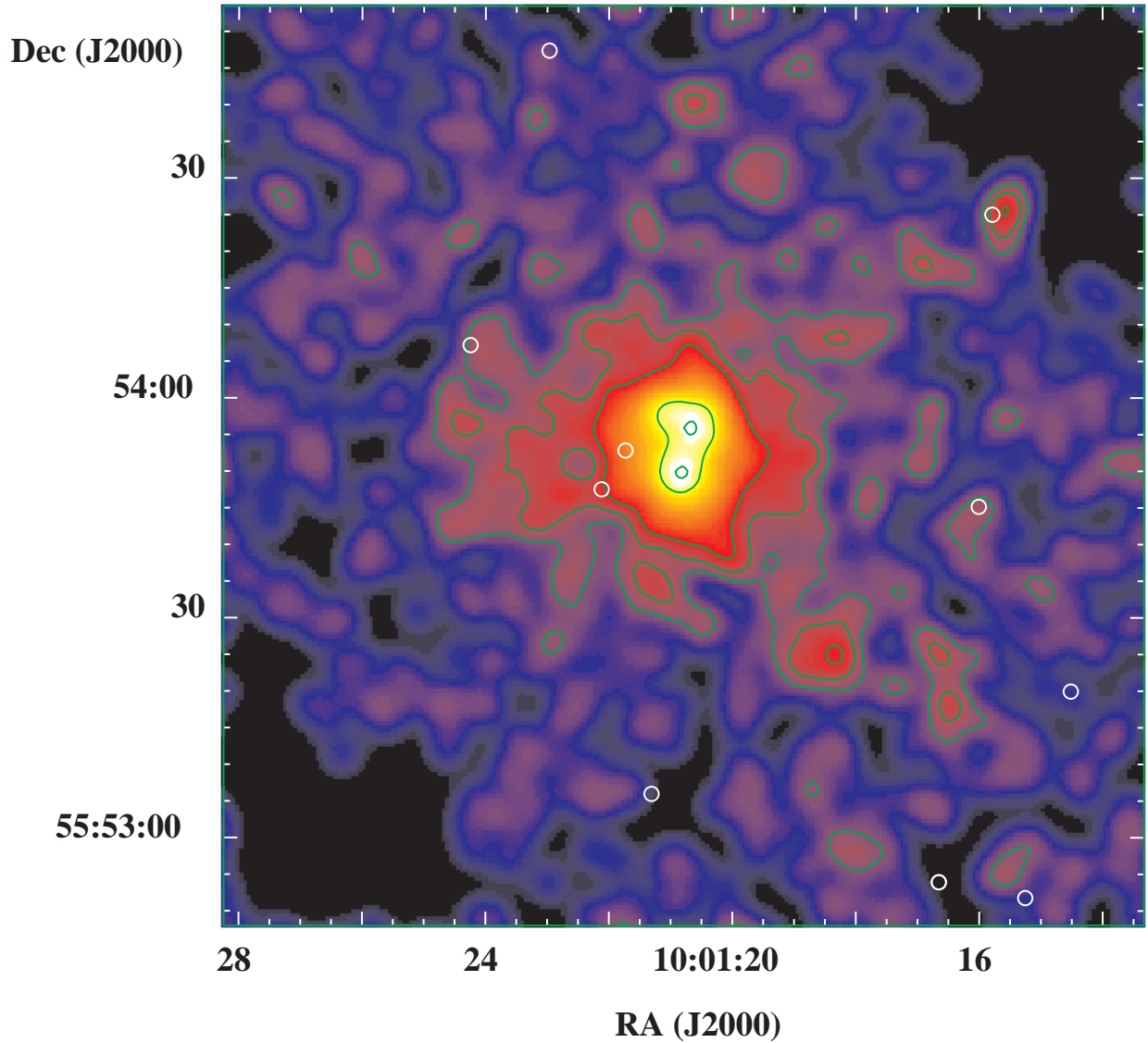}{8.in}{0}{120.}{120.}{-350}{-120}
\protect\caption
{\small An adaptively smoothed X-ray image of the gravitational
lens system Q0957+561. To improve the signal-to-noise of the lens cluster
we selected X-rays with energies ranging between 0.5 and 3~keV.
The lens cluster is resolved and appears to be centered
slightly northeast of image B. The contour levels are  0.1, 3.5$\times$ 10$^{-4}$,
3.5$\times$ 10$^{-5}$, 2.2$\times$ 10$^{-5}$, and 1.3$\times$ 10$^{-5}$ of the peak
emission of image A. The white circles indicate the optical locations of
the galaxies in the cluster at z = 0.36. Galaxy positions are taken
from Angonin-Willaime et al. (1994). North is up east is left.
\label{fig:fig1}}
\end{figure*}

\clearpage
\begin{figure*}[t]
\plotfiddle{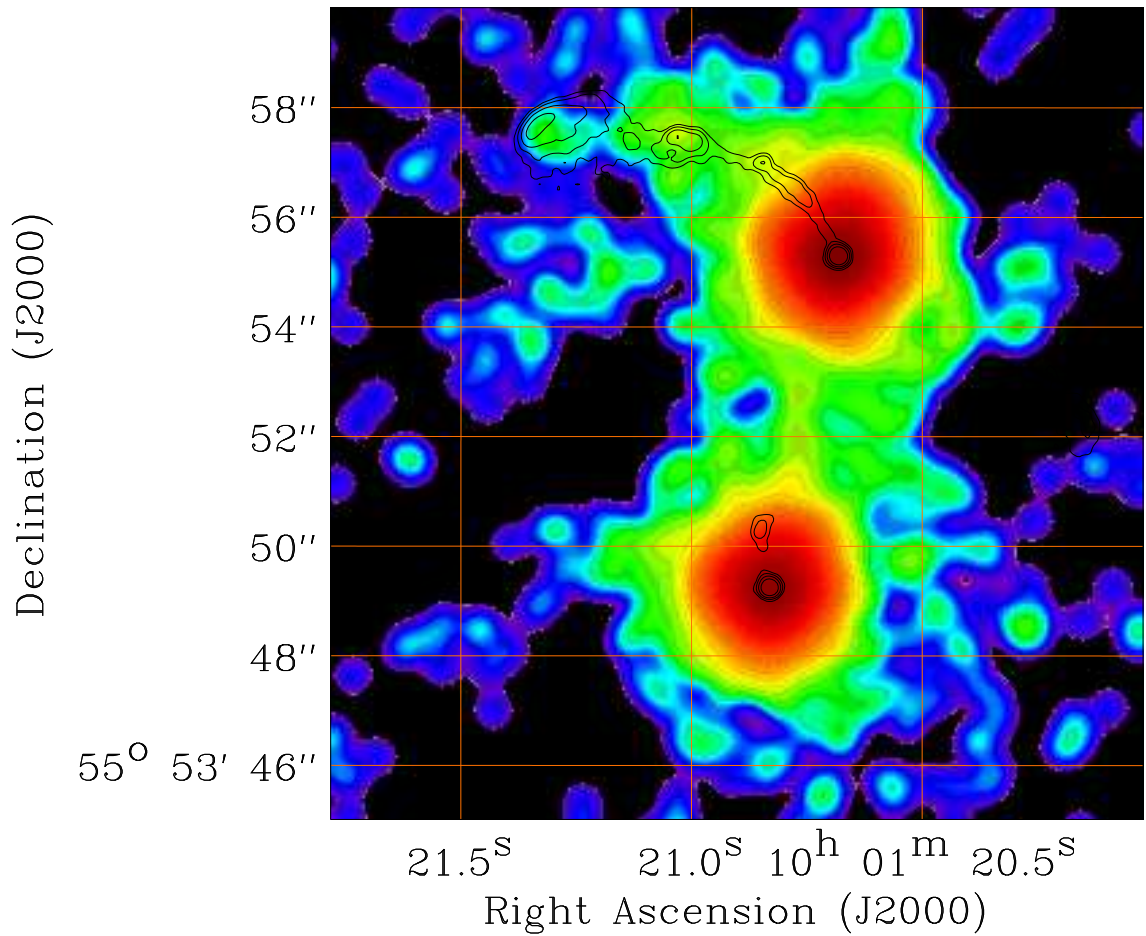}{6.5in}{0}{100.}{100.}{-280}{-130}
\protect\caption
{\small An overlay of the VLA and {\sl Chandra} images of Q0957+561.
The contours represent the radio 3.6~cm VLA image of Q0957+561 
provided courtesy of Harvanek et al. (1997). 
The X-ray and radio data are binned with a bin size of 0{\sarc}048 on a side. 
For presentation purposes the X-ray data are smoothed with a 
circular Gaussian of radius $\sigma$ = 0{\sarc}25 resulting in an effective 
resolution of $\sigma$ $\sim$ 0{\sarc}33. The overlay clearly 
shows that the shape and angular structure of the X-ray and 
radio jets are similar. 
The radio contour levels are 0.11, 3.5$\times$ 10$^{-2}$, 1.1$\times$ 10$^{-2}$
and 3.5$\times$ 10$^{-3}$ of the peak emission of image A. The beam width
is 0{\sarc}24  $\times$ 0{\sarc}17. The X-ray image
contains photons with energies ranging between 0.5 and 10~keV.
The brightness profiles of the X-ray and radio jets differ significantly
as shown more clearly in Figure 3. North is up east is left.
\label{fig:fig2}}
\end{figure*}

\clearpage
\begin{figure*}[t]
\plotfiddle{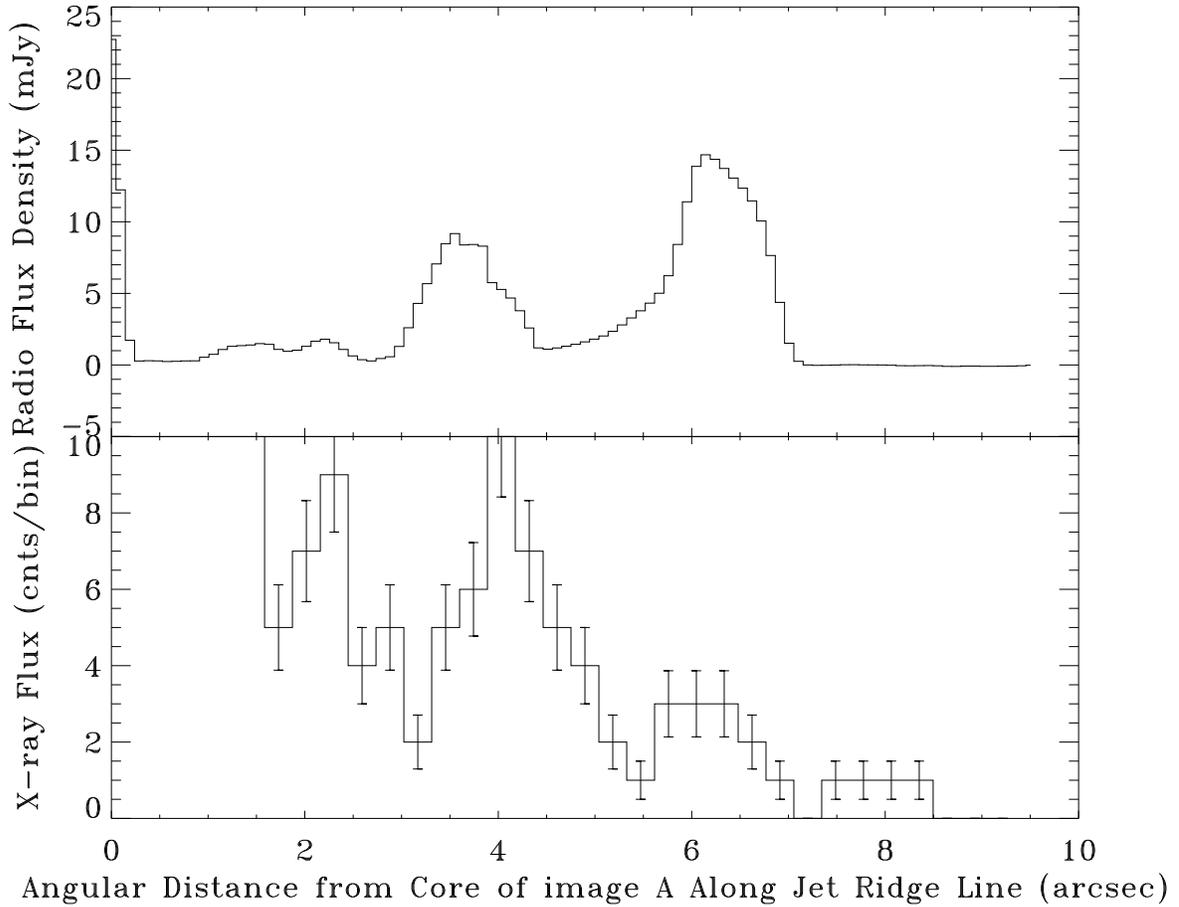}{6.5in}{0}{100.}{100.}{-320}{-250}
\protect\caption
{\small The radio and X-ray intensity profiles 
are plotted as a function of angular distance from the core
of image A along the jet ridge line.
The X-ray profile provides counts in 0{\sarc}288
increments integrated $\pm$ 0{\sarc}48 perpendicular to the
jet ridge line and the radio profile provides the 3.6~cm flux density in 
0{\sarc}096 increments integrated $\pm$ 0{\sarc}48 perpendicular to the
jet ridge line. Radio counterparts within 0{\sarc}5
of the X-ray bright knots 
located 4$''$ and 6$''$ from the core are clearly visible. 
Whereas the angular structure of the X-ray and radio jets are similar, the 
intensity distributions along the jet appear to be anti-correlated. 
In particular, the X-ray brightness decreases along the jet,
whereas the radio flux density of the knots increases 
with distance from the core (see text for interpretation). 
\label{fig:fig3}}
\end{figure*}

\clearpage
\begin{figure*}[t]
\plotfiddle{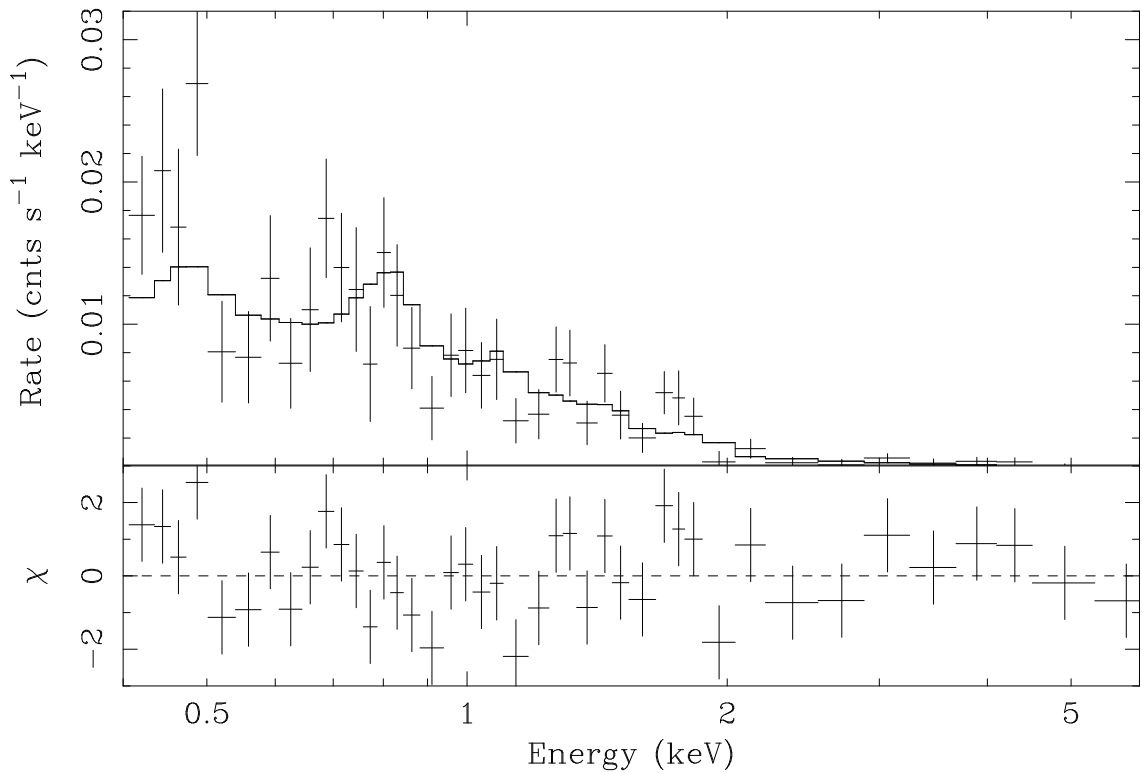}{7.5in}{0}{100.}{100.}{-290}{-300}
\protect\caption
{\small (Top panel) The X-ray spectrum of the lens cluster
along with the best-fit thermal Raymond-Smith model (see Table 2). 
The model is consistent with a plasma temperature of $\sim$ 2.1~keV.
The detected spectral feature between 0.7 and 0.9~keV (observed-frame) 
correspond to a redshifted z = 0.36 complex of Fe~L lines,
confirming the detection of the lens cluster.
The uncertainty in the calibration of ACIS S3 below 0.5~keV
contributes to the large residuals between 0.4 and 0.5~keV.
(Lower panel) Residuals in units of standard deviations
with error bars of size 1$\sigma$.
\label{fig:fig4}}
\end{figure*}

\clearpage
\begin{figure*}[t]
\plotfiddle{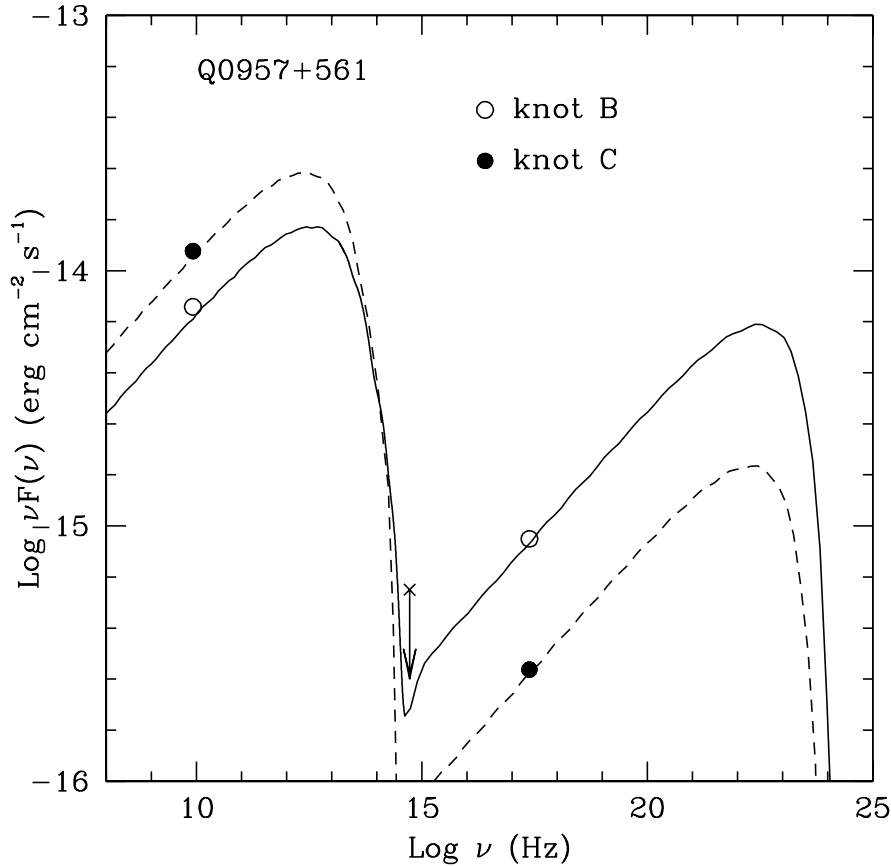}{6.5in}{0}{60.}{60.}{-180}{0}
\protect\caption
{\small The spectral energy distributions (SEDs) of jet knots
B (open circles) and C (filled circles). The radio flux densities were extracted from the 
radio $\lambda$3.6~cm VLA data kindly provided by 
Michael Harvanek of Apache Point Observatory.  
The optical upper limit (cross) is based on published 
results from a deep HST observation of 
the field (Bernstein \& Fischer 1997). 
The values of ${\nu}F_{\nu}$ of the jet knots in the X-ray band are 
at least an order of magnitude above an 
extrapolation of the radio and optical 
data points, suggesting that the emission 
in the radio, optical and X-ray 
bands is not consistent with a single synchrotron model 
with a single power law distribution. 
The solid and dashed lines represent fits
of the CMB model (see text) to the spectra of knots B and C, 
respectively. The CMB model, in which cosmic microwave photons are inverse
Compton scattered by synchrotron-emitting electrons in the jet, has the following parameters.
{\sl Knot B:} The emission region is assumed to be spherical with a 
radius of $R$ = 1 $\times$ 10$^{23}$~cm, a magnetic field 
intensity of $B$ = 6 $\times$ 10$^{-5}$~G, and a Doppler factor 
of $\delta$ = 1.4. The electron distribution is assumed to be 
a power-law with extremes of $\gamma_{min}$ = 10 and $\gamma_{max}$ = 3 $\times$ 10$^{5}$, 
a slope of n = 2.6, and a normalization
of K = 6.4 $\times$ 10$^{-5}$ cm$^{-3}$.
{\sl Knot C:} The CMB model for knot C has the same parameters as knot B 
except K = 4 $\times$ 10$^{-4}$ cm$^{-3}$ 
and $\delta$ = 1.05.
\label{fig:fig5}}
\end{figure*}

\clearpage
\begin{figure*}[t]
\plotfiddle{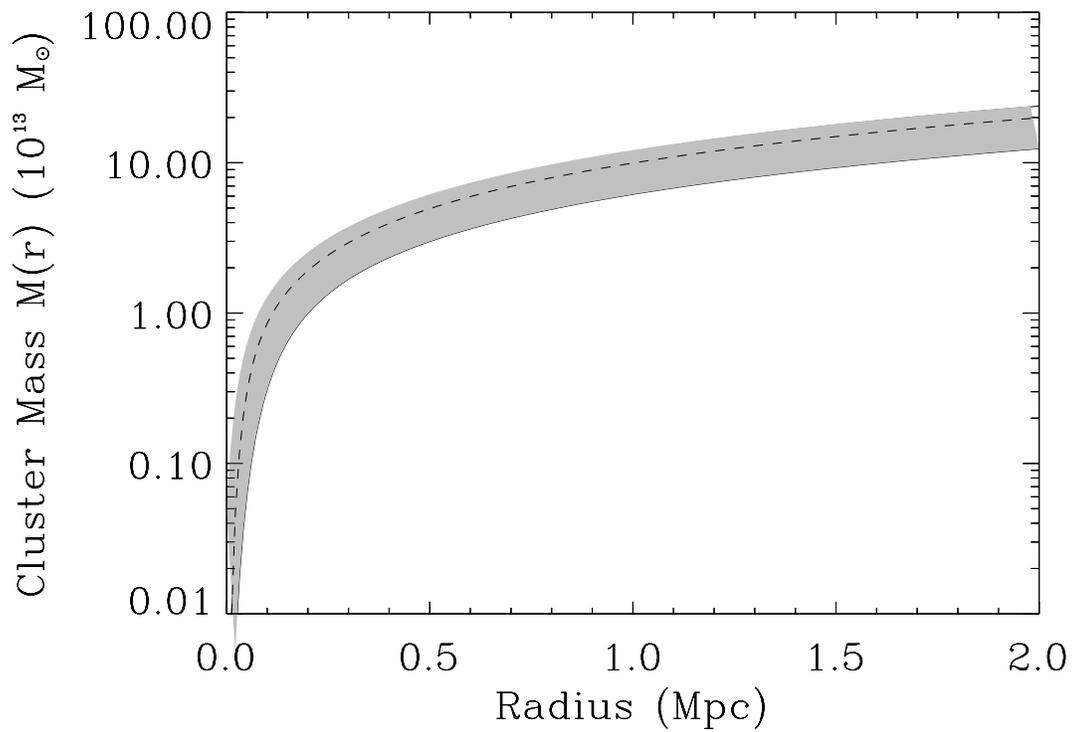}{7.5in}{0}{90.}{90.}{-270}{-200}
\protect\caption
{\small The total mass of the lens cluster within
a radius $r$. The dashed line corresponds to the best-fit spatial
and spectral parameters. The shaded region indicates the allowable 
range of the cluster mass for the estimated uncertainties
of the best-fit spatial and spectral parameters.
\label{fig:fig6}}
\end{figure*}

\clearpage
\begin{figure*}[t]
\plotfiddle{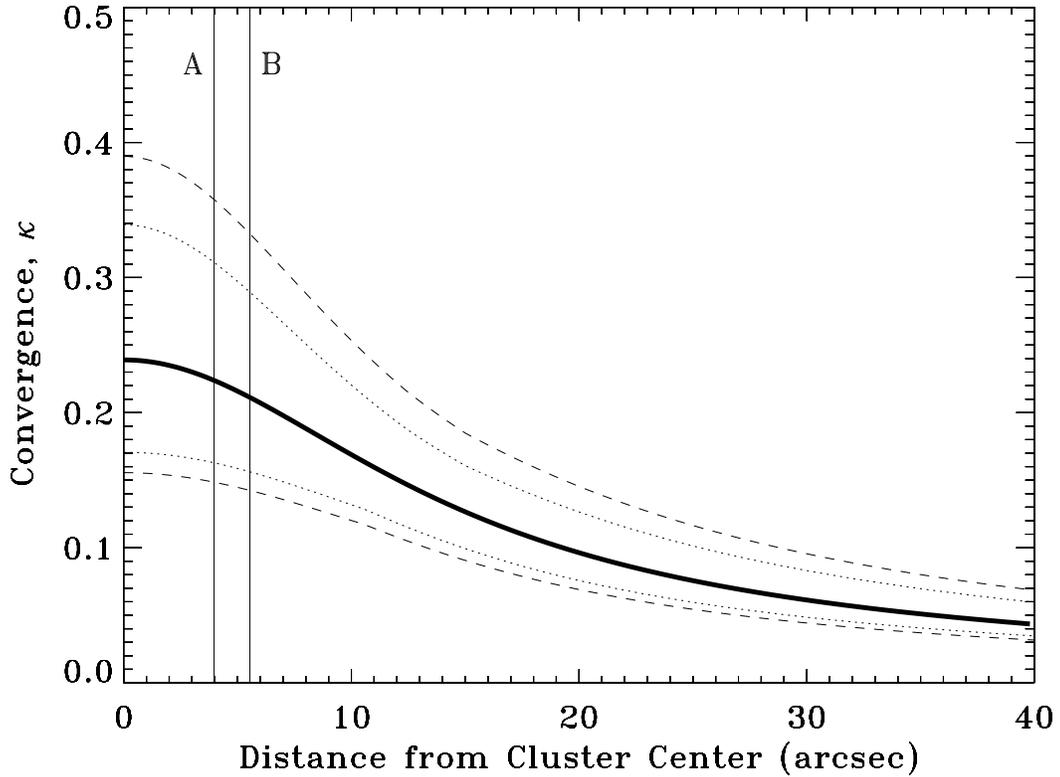}{7.5in}{0}{90.}{90.}{-290}{-200}
\protect\caption
{\small The convergence parameter $\kappa$ of the
lens cluster as a function of distance from the cluster center.
The thick solid line corresponds to the best-fit spatial and spectral parameters.
The largest contributor to the uncertainty in the present 
measurement of $\kappa(x)$ is the weak constraint on the temperature of the cluster.
To illustrate this weakness we have plotted the uncertainty in
$\kappa(x)$ assuming 68\% (dotted lines) and 90\% (dashed lines) 
confidence intervals for the temperature. 
We also chose cluster limits ranging from 0.7$r_{500}$ and 1.4$r_{500}$,
where $r_{500}$ is the radius in which the mean over-density is 500,
and $r_{500}$ = 1.58 $h^{-1}_{75}$ Mpc (T$_{X}$/10~keV)$^{1/2}$ $\sim$ 0.71 $h^{-1}_{75}$ Mpc
(Mohr, Mathiesen, \& Evrard, 1999). The solid vertical lines indicate the 
distances of images A and B from the cluster center.
\label{fig:fig7}}
\end{figure*}

\end{document}